\documentstyle[prl,aps,eqsecnum,floats,epsf]{revtex}
\begin{document}
\title{Coulomb drag in longitudinal
magnetic field in quantum wells
}
\author{V. L. Gurevich and M. I. Muradov}
\address{Solid State Physics Division, A.F.Ioffe Institute,
194021 Saint Petersburg, Russia}
\date{\today}
\maketitle
\begin{abstract}
\baselineskip=3.5ex
The influence of a longitudinal magnetic field on the Coulomb drag
current created in the ballistic transport regime in a quantum well
by a ballistic current in a nearby parallel quantum well is
investigated. We consider the case where the magnetic field is so
strong that the Larmour radius is smaller than the width of the
well. Both in Ohmic and non-Ohmic case, sharp oscillations of
the drag current as a function of the gate voltage or chemical
potential are predicted. We also study dependence of the drag current
on the voltage $V$ across the driving wire, as well as on the
magnetic field $B$.

Studying the Coulomb drag one can make conclusions about the electron
spectrum and and electron-electron interaction in quantum wells.
\end{abstract}
\baselineskip=3.5ex
\section{Introduction}
The influence of a magnetic field on the Coulomb drag are
investigated in different geometries. The Coulomb drag between two
two-dimensional (2D) quantum wells in a strong magnetic field
perpendicular to the planes of the wells and in the presence of
disorder has been investigated in Ref.~\cite{Bonsager}. In magnetic
field perpendicular to the planes the Hall voltage can be induced in
the drag quantum well in the direction perpendicular to both
direction of the magnetic field and of the current in the drive
well~\cite{Kamenev},~\cite{Hu}. These two geometries can be called
transverse.

The purpose of the present paper is to study the influence of an
in-well magnetic field $\bf B$ on the Coulomb drag current in the
course of ballistic (collisionless) electron transport in a quantum
well due to a ballistic drive current in a parallel quantum well. In
other words, we consider the longitudinal geometry, i.e. the case
where the magnetic field is parallel to the applied electric field
$\bf E$ and to the plane of the well itself.

We will concern ourselves with the case of a strong magnetic
field that makes the motion of the carriers along the field one
dimensional and alters the density of electron states.
Moreover, we restrict ourselves with the quantum limit when only
the ground Landau oscillator states are occupied by electrons in
the two quantum wells, so that
\begin{equation}
\hbar\omega_B\gtrsim\mu
\label{1z}
\end{equation}
Here $\omega_B$ is the cyclotron frequency while $\mu$ is the
chemical potential. A theory of electronic transport through
three-dimensional ballistic microwires in longitudinal magnetic
fields at low temperatures has been developed in
Ref.~\cite{Bogachek}. Our geometry is similar to that considered
in Ref.~\cite{Bogachek}. However, in the present paper we
consider much simpler situation of a very strong magnetic field
satisfying Eq. (\ref{1z}). Later on we hope to return to a more
general case of a weaker magnetic field where several Landau
levels may be involved.

The magnetic field making the motion of the electrons in the
transverse direction one dimensional maps the problem under
consideration onto the Coulomb drag problem in two
one-dimensional wires already considered by the authors in
Ref.~\cite{GurMur} in the Fermi liquid approach. Therefore, our
final formulae for the Coulomb drag current appear to be similar
to those obtained in~\cite{GurMur}. Physically the magnetic
field may play the following important role. It will suppress
the tunneling of electrons between the quantum wells that, if
present, would impede observation of the Coulomb drag.

The magnetic field may change the electron quasimomentum
relaxation time. Scattering of electrons by ionized impurities in
sufficiently strong magnetic fields may be even weaker than for
$B=0$. As for the relaxation due to the phonon scattering, a strong
magnetic field can alter the density of electron states and in
the quantum limit the relaxation rate may be bigger than for
$B=0$. We, however, will assume the temperature to be so low
that the transport remains ballistic even in the presence of
magnetic field.

We consider the case where the magnetic length 
\begin{equation}\label{2z}
a_B=\sqrt{
{\hbar\,c\over{|e|B}} } 
\end{equation} 
is much smaller than the distance $W$
between the quantum wells of the width $L_x\sim\,W$ each
\begin{equation}\label{ineq_10}
W/{a_B}\gg\,1.
\end{equation}

This inequality establishes lower bound for the values of the
magnetic field for a given distance between the quantum wells. For
instance, for $W\sim\,80\,\,$nm the inequality requires magnetic
fields of the order of $B\sim\,1\,\,$T, or bigger.

It is convenient to break our calculations into several parts.  In
the first part we will give the principal equations of our theory
based on the Boltzmann treatment of the transport. We will consider a
linear response in Sec.~\ref{lr}. Next we will discuss a non-Ohmic
case in Sec.~\ref{nohm}. Comparison of our results with the 1D
Coulomb drag results in the longitudinal geometry and 2D Coulomb drag
results for $B=0$ will be given in Summary.

\section{Boltzmann equation}\label{be}

We consider two parallel quantum wells perpendicular to $x$ axis.  The
eigenfunctions and eigenvalues for a one electron problem in a
magnetic field along the $z$ axis in the $i$th quantum well is [we use
the gauge ${\bf A}=(0,Bx,0)$]
\begin{equation}\label{bol_10}
\psi_{0p_yp_z}={1\over{\sqrt{L_yL_z}}}
\varphi_0\left({x-x_{p_y}\over{a_B}}\right)\exp(ip_yy/\hbar+ip_zz/\hbar),
\end{equation}
\begin{equation}\label{bol_11}
\varepsilon_{0p_z}=U_i+{\hbar\omega_B\over{2}}+{p_z^2\over{2m}}.
\end{equation}
Here $m$ is the effective electron mass,
$\varphi_0\left[(x-x_{p_y})/a_B\right]$ is the wave function of
a harmonic oscillator in the ground state oscillating about the
point $x_{p_y}=-a_B^2p_y/\hbar=-p_y/m\omega_B$.
The wave function $\psi_{0p_yp_z}$
describes a state for which the electron probability
distribution is large only within the slab of the width $\approx
a_B$ symmetrically situated about the plane $x=x_{p_y}$ and
falls off exponentially outside the slab. As we consider the
case $L_x\,\gg\,a_B$ we will assume the wave function to be
equal $\psi_{0p_yp_z}$ if $x_{p_y}$ is within the quantum well
and zero otherwise.  In what follows we will need the matrix
elements of the functions $\exp(\pm\,i{\bf qr})$ between two
stationary states. We have
\begin{eqnarray}\label{bol_12}
\langle0p^{\prime}_yp^{\prime}_z|e^{\pm\,i{\bf qr}}|
0p_yp_z\rangle=e^{\pm\,iq_x(x_{p_y}+x_{p^{\prime}_y})/2}
\nonumber\\
\times e^{-a_B^2q_x^2/4}e^{-(x_{p_y}-x_{p^{\prime}_y})^2/4a_B^2}
\delta_{p^{\prime}_z,p_z\pm\hbar\,q_z}
\delta_{p^{\prime}_y,p_y\pm\hbar\,q_y}.
\end{eqnarray}

The diagram representing Coulomb drag effect is illustrated in Fig.~1.
\begin{figure}[htb]
\epsfxsize=4.4in
\epsffile{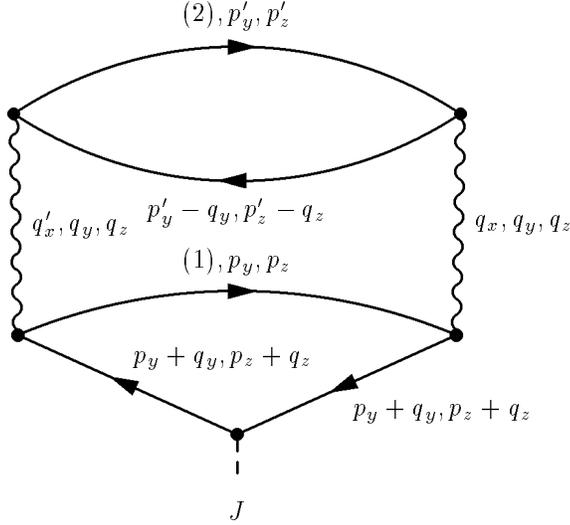}
\caption{Coulomb drag diagram. Here the labels 2, (1) stand for the
drive (drag) quantum wells.}
\label{fig1}
\end{figure} 
The
external driving force enters the diagram through nonequilibrium
distribution function represented by the solid lines marked by the
symbol 2 indicating that they represent the drive quantum well.

Now we embark on analysis of the conservation laws for the collisions
of electrons belonging to two different quantum wells. We have
\begin{equation}\label{cl_1}
\varepsilon_{0p_z}^{(1)}+\varepsilon_{0p_z^{\prime}}^{(2)}=
\varepsilon_{0p_z+\hbar\,q_z}^{(1)}+
\varepsilon_{0p_z^{\prime}-\hbar\,q_z}^{(2)}
\end{equation}
where
$\varepsilon_{0p_z}^{(1,2)}=U_{1,2}+\hbar\omega_B/2+p_z^2/2m$.

The solution of Eq.(\ref{cl_1})
is
\begin{equation}\label{solut}
\hbar\,q_z=p_z^{\prime}-p_z.
\end{equation}
The $\delta-$function describing energy conservation can
be recast into the form
\begin{equation}\label{simp}
\delta(\varepsilon_{np_z}^{(1)}+\varepsilon_{lp_z^{\prime}}^{(2)}-
\varepsilon_{np_z+\hbar\,q_z}^{(1)}-\varepsilon_{lp_z^{\prime}-
\hbar\,q_z}^{(2)})=
{m\over{\hbar\,|q_z|}}\delta[\hbar\,q_z-(p_z^{\prime}-p_z)].
\end{equation}
Therefore, the initial quasimomenta $p_z$ and $p_z^{\prime}$ after
the collision become $p_z+\hbar\,q_z=p_z^{\prime}$ and
$p_z^{\prime}-\hbar\,q_z=p_z$, i.e. the electrons swap their
quasimomenta as a result of collision.

Following~\cite{GurMur,GM} we assume that the drag current in the
quantum well $1$ is much smaller than the drive ballistic
current in the quantum well $2$ and calculate it by solving the
Boltzmann equation for the quantum well $1$.  We have
\begin{equation}\label{df}
v_z{\partial \Delta F_{0p_y}^{(1)}(p_z,z)\over{\partial z}}=
-I^{(12)}\{F^{(1)},F^{(2)}\},
\end{equation}
where $F^{(1,2)}$ are the electron distribution functions in the
quantum wells $1$ and $2$ respectively, and the collision integral
$I^{(12)}\{F^{(1)},F^{(2)}\}$ takes
into account the interwell electron-electron scattering
\begin{equation}\label{ci}
I^{(12)}\{F^{(1)},F^{(2)}\}=\sum_{p_z^{\prime}p^{\prime}_yq^{\prime}_x
{\bf q}}
W_{1p_zn,2p_z^{\prime}}^{1p_z+\hbar q_z,2p_z^{\prime}-\hbar q_z}
(q^{\prime}_x,q_x,q_y,p_y,p_y^{\prime}){\cal S}.
\end{equation}
In this expression the sum over $p^{\prime}_y$ should be determined
by the requirement that the $x$-center of the oscillator function
is within the second quantum well. The requirement imposes the
constraint
\begin{equation}\label{constraint}
{\hbar\over{a_B^2}}\left(W+L_x-{L_x\over{2}}\right)\,<\,p^{\prime}_y\,<\,
{\hbar\over{a_B^2}}\left(W+L_x+{L_x\over{2}}\right),
\end{equation}
and the product of distribution functions $\cal S$ is
\begin{equation}\label{cS}
{\cal S}=F^{(1)}_{0p_z}F^{(2)}_{0p_z^{\prime}}
\left(1-F^{(1)}_{0p_z+\hbar q_z}\right)\left(1-F^{(2)}_{0p_z^{\prime}-
\hbar q_z}\right)
-F^{(1)}_{0p_z+\hbar q_z}F^{(2)}_{0p_z^{\prime}-\hbar q_z}
\left(1-F^{(1)}_{0p_z}\right)\left(1-F^{(2)}_{0p_z^{\prime}}\right).
\end{equation}
\begin{equation}\label{dist_fun}
F^{(1)}_{0p_z}=\theta[v_z]f(\varepsilon_{0p_z}-\mu_B^{1L})+
\theta[-v_z]f(\varepsilon_{0p_z}-\mu_B^{1R})+\Delta F^{(1)}_{0p_z}
\end{equation}
where
$$
\theta[v_z]=\left\{{1\quad\mbox{for}\quad v_z>0\atop 0\quad\mbox{for}
\quad v_z<0}\right..
$$ Here we assume that the electrons move ballistically within
the quantum well and the electrons moving from the left and
right reservoirs have the chemical potentials
$\mu_B^{1L}=\mu_B-eV_{\mbox{d}}/2$ and
$\mu_B^{1R}=\mu_B+eV_{\mbox{d}}/2$ respectively.
We introduce also the drag voltage $V_{\mbox{d}}$
induced across the drag quantum well due to the quasimomentum transfer
from the driving quantum well, i.e. we assume an open circuit
for the drag quantum well.

The solution of Eq.(\ref{df}) is (here we omit the equilibrium part)
\begin{equation}\label{df_1}
\Delta F^{(1)}_{0p_z}=-\left(z\pm {L\over{2}}\right){1\over{v_z}}
I^{(12)}\{F^{(1)},F^{(2)}\},\quad
\mbox{for}\quad {p_z\,>\,0,\atop p_z\,<\,0.}
\end{equation}
Using the particle conserving
property of the scattering integral
\begin{equation}\label{pcp}
\sum_{p_zp_y}I^{(12)}\{F^{(1)},F^{(2)}\}=0
\end{equation}
we get for the total current in the drag quantum well defined as
\begin{equation}\label{curt}
J={e\over
L_z}\sum_{p_yp_z}v_zF^{(1)}_{0p_z},
\end{equation}
the result
\begin{equation}\label{current}
J=-e\sum_{p_y,(p_z>0)}I^{(12)}\{F^{(1)},F^{(2)}\}+e{1\over
L_z}\sum_{p_y,(p_z>0)}v_z
[f(\varepsilon_{0p_z}-\mu_B^{1L})-f(\varepsilon_{0p_z}-\mu_B^{1R})].
\end{equation}
In these equations the sum over $p_y$ is restricted by the requirement
that the $x$-center of Landau oscillator must be within the
quantum well, so that
$-\hbar\,L_x/2a_B^2\,<\,p_y\,<\,\hbar\,L_x/2a_B^2$.  Introducing the
density of states (including spin) per unit quasimomentum interval
\begin{equation}\label{dos}
N(p_z)dp_z=2{L_z\over{(2\pi\hbar)^2}}{\hbar\,L_xL_y\over{a_B^2}}dp_z
\end{equation}
we have
\begin{equation}\label{dir_current}
J_{\mbox{Ohm}}=
-e{2eV_{\mbox{d}}\over{(2\pi\hbar)^2}}{\hbar\,L_xL_y\over{a_B^2}}
\int_{U_1+\hbar\omega_B/2}^{\infty}\,d\varepsilon\,
\left(-{\partial f(\varepsilon-\mu_B)\over{\partial \varepsilon}}\right).
\end{equation}
For the degenerate electron gas this expression can be written as
\begin{equation}\label{dir_c}
J_{\mbox{Ohm}}=
-{e^2\over{\pi\hbar}}{L_xL_y\over{2\pi a_B^2}}V_{\mbox{d}}.
\end{equation}
Here the number of Larmour circles covering the cross section of
the quantum well $L_xL_y/2\pi a_B^2$ appears instead of the
number of open channels in the 1D situation.

We assume that only the ground Landau oscillator state is occupied,
so that
\begin{equation}\label{dir_cad}
U_1+{1\over{2}}\hbar\omega_B<\,\mu_B\,<\,U_1+{3\over{2}}\hbar\omega_B.
\end{equation}
Taking into account Eq.(\ref{simp}) we obtain for the Coulomb
scattering probability Eq.(\ref{ci})
\begin{eqnarray}\label{sc_pr}
W_{1p_z,2p_z^{\prime}}^{1p_z+\hbar q_z,2p_z^{\prime}-\hbar q_z}
(q^{\prime}_x,q_x,q_y,p_y,
p_y^{\prime})={m\over{\hbar |q_z|}}\delta[\hbar q_z-(p_z^{\prime}-p_z)]
{2\pi\over{\hbar}}U_{{\bf q}}U_{q_x^{\prime}q_yq_z}\nonumber\\
\times\langle0p_zp_y|e^{-iq_x^{\prime}x-iq_yy-iq_zz}|0p_z+\hbar
q_zp_y+ \hbar q_y\rangle
\langle0p_z+\hbar q_zp_y+\hbar
q_y|e^{iq_xx+iq_yy+iq_zz}|0p_zp_y\rangle \nonumber\\
\times\langle0p^{\prime}_zp^{\prime}_y|e^{iq_x^{\prime}x+iq_yy+iq_zz}|0
p^{\prime}_z-\hbar q_zp^{\prime}_y-\hbar q_y\rangle
\langle0p^{\prime}_z-\hbar q_zp^{\prime}_y-\hbar q_y
|e^{-iq_xx-iq_yy-iq_zz}
|0p^{\prime}_zp^{\prime}_y\rangle.
\end{eqnarray}
Here we use the unscreened Coulomb potential postponing
discussion as to when this approximation can be justified until the
last section.

To calculate the drag current we iterate the Boltzmann equation
in the interwell collision term that we assume to be small.
Therefore one can choose the distribution functions in the
collision term to be equilibrium ones, e.g.
$F^{(1)}_{0p}=f(\varepsilon^{(1)}_{0p}-\mu_B)$ for the
first quantum well.

We assume, in the spirit of the approach developed by
Landauer~\cite{Lan}, Imry~\cite{LB} and B\"uttiker~\cite{B} the drive
quantum well to be connected to reservoirs which we call 'left` $l$
and 'right` $r$.  Each of them is in independent equilibrium
described by the shifted chemical potentials $\mu_B^l=\mu_B-eV/2$ and
$\mu_B^r=\mu_B+eV/2$, where $\mu_B$ is the equilibrium chemical
potential in the magnetic field. Therefore, the electrons entering
quantum well from the 'left`('right`) and having quasimomenta
$p_z^{\prime}>0$ ($p_z^{\prime}<0$) are described by
$F^{(2)}_{0p_z^{\prime}}=
f(\varepsilon^{(2)}_{0p_z^{\prime}}-\mu_B^l)$
[$F^{(2)}_{0p_z^{\prime}}=
f(\varepsilon^{(2)}_{0p_z^{\prime}}-\mu_B^r)$] and we see that the
collision integral Eq.(\ref{ci}) is identically zero if the initial
$p_z^{\prime}$ and final $p_z^{\prime}-q$ quasimomenta in the drive
quantum well are of the same sign. This means that only the
{\em backscattering processes} contribute to the drag current.

Due to Eq.(\ref{simp}) we are left only with
$p_z^{\prime}\,<\,0$ (since we are restricted according to
Eq.(\ref{current}) by the constraint $p_z^{\prime}-\hbar q_z=p_z\,>0$)
and obtain in view of the $\delta$-function in
Eq.(\ref{sc_pr}) the following product of distribution
functions in the collision term
\begin{equation}\label{ct}
{\cal P}=F^{(1)}_{0p_z}F^{(2)r}_{0p_z^{\prime}}
\left(1-F^{(1)}_{0p_z^{\prime}}\right)\left(1-F^{(2)l}_{0p_z}\right)
-F^{(1)}_{0p_z^{\prime}}F^{(2)l}_{0p_z}
\left(1-F^{(1)}_{0p_z}\right)\left(1-F^{(2)r}_{0p_z^{\prime}}\right),
\end{equation}
or
\begin{eqnarray}\label{ct_1}
{\cal P}=f(\varepsilon^{(1)}_{0p_z}-\mu_B)
f(\varepsilon^{(2)}_{0p_z^{\prime}}-\mu_B^r)
[1-f(\varepsilon^{(1)}_{0p_z^{\prime}}-\mu_B)
][1-f(\varepsilon^{(2)}_{0p_z}-\mu_B^l)]\nonumber\\
-f(\varepsilon^{(1)}_{0p_z^{\prime}}-\mu_B)f(\varepsilon^{(2)}_{0p_z}-\mu_B^l)
[1-f(\varepsilon^{(1)}_{0p_z}-\mu_B)]
[1-f(\varepsilon^{(2)}_{0p_z^{\prime}}-\mu_B^r)].
\end{eqnarray}
This equation will be analyzed in the following sections.
\section{Linear response}\label{lr}
In this case $eV/T\,\ll\,1$ (we assume the
Boltzmann constant to be equal 1) and Eq.(\ref{ct_1}) can be
recast into the form
\begin{equation}\label{ct_02}
{\cal P}={eV\over{T}}f(\varepsilon^{(1)}_{0p_z}-\mu_B)
f(\varepsilon^{(2)}_{0p_z^{\prime}}-\mu_B)
[1-f(\varepsilon^{(1)}_{0p_z^{\prime}}-\mu_B) ]
[1-f(\varepsilon^{(2)}_{0p_z}-\mu_B)].
\end{equation}

Shifting the integration variable
$p^{\prime}_y\rightarrow\,p^{\prime}_y+\hbar\,(W+L_x)/a_B^2$
we have for the drag current
\begin{eqnarray}\label{drag_cur}
J_{\mbox{drag}}=-e{eV\over{T}}{2\pi\over{\hbar}}
\left({4\pi\,e^2\over{\kappa}}\right)^2{1\over{2\pi\hbar}}
\int_0^{\infty}{2L_zdp_z\over{2\pi\hbar}}\int_0^{\infty}
{dp_z^{\prime}\over{2\pi\hbar}}{m\over{(p_z+p_z^{\prime})}}\nonumber\\
\times f(\varepsilon^{(1)}_{0p_z}-\mu_B)
f(\varepsilon^{(2)}_{0p_z^{\prime}}-\mu_B)
[1-f(\varepsilon^{(1)}_{0p_z^{\prime}}-\mu_B) ]
[1-f(\varepsilon^{(2)}_{0p_z}-\mu_B)]\nonumber\\
\times
\int_{-\hbar\,L_x/2a_B^2}^{\hbar\,L_x/2a_B^2}{2L_ydp_y\over{2\pi\hbar}}
{dp_y^{\prime}\over{2\pi\hbar}}
g_{00}\left[(p_z+p_z^{\prime})/\hbar,(p_y-p_y^{\prime})/\hbar\right]
\end{eqnarray}
where
\begin{eqnarray}\label{vz}
g_{00}(k_z,k_y)=\int_{-\infty}^{\infty}{dq_y\over{2\pi}}
e^{-a_B^2q_y^2}A^2(k_z,k_y,q_y),
\end{eqnarray}
\begin{eqnarray}\label{vz01}
A(k_z,k_y,q_y)=\int_{-\infty}^{\infty}
{dq_x\over{2\pi}}
{e^{-ia_B^2q_x[k_y-(W+L_x)/a_B^2+q_y]}
e^{-a_B^2q_x^2/2}
\over{(k_z^2+q_{\perp}^2) }}
\end{eqnarray}
$q_{\perp}^2=q_x^2+q_y^2$.
The last integral over the centers of the Larmour circles (since
$x_p=-a_B^2p_y/\hbar$) in Eq.(\ref{drag_cur}) plays the role of
an effective Coulomb interaction potential between the electrons freely
moving along the direction of applied magnetic field.
\begin{eqnarray}
\int\limits_{-\hbar\,L_x/2a_B^2}^{\hbar\,L_x/2a_B^2}{2L_ydp_y\over{2\pi\hbar}}
{dp_y^{\prime}\over{2\pi\hbar}}
g_{00}\left[{p_z+p_z^{\prime}\over\hbar},
{p_y-p_y^{\prime}\over\hbar}\right]\nonumber\\
={2L_y\over{(2\pi)^2}}\int\limits_{0}^{L_x/a_B^2}dk_y\,k_y
\left\{g_{00}\left[{p_z+p_z^{\prime}\over\hbar},{L_x\over a_B^2}-k_y\right]+
g_{00}\left[{p_z+p_z^{\prime}\over\hbar},k_y-{L_x\over a_B^2}\right]
\right\}
\end{eqnarray}
We keep only the first term in this expression since the second
term includes a faster oscillating exponent
$\sim\,\exp{(iq_x(W+L_x))}$ as compared to the oscillating
exponent in the first term $\sim\,\exp{(iq_xW)}$.

As ${W/{a_B}}\gg1$ we can sufficiently simplify the expression
for $g_{00}$. We obtain
\begin{equation}
g_{00}(k_z,k_y)=e^{a_B^2k_z^2}\int_{-\infty}^{\infty}{dq_y\over{2\pi}}
A^2(k_z,k_y,q_y).
\end{equation}
\begin{eqnarray}
A(k_z,k_y,q_y)\simeq\,\int\,{dq_x\over{2\pi}}
{e^{iq_x[(W+L_x)-a_B^2k_y]}\over{q_x^2+q_y^2+k_z^2}}
={e^{-|W+L_x-a_B^2k_y|\sqrt{q_y^2+k_z^2}}
\over{2\sqrt{q_y^2+k_z^2}}}
\end{eqnarray}
Finally, the interaction term acquires the form
\begin{eqnarray}\label{fin_form}
\int_{-\hbar\,L_x/2a_B^2}^{\hbar\,L_x/2a_B^2}{2L_ydp_y\over{2\pi\hbar}}
{dp_y^{\prime}\over{2\pi\hbar}}
g_{00}\left[k_z,(p_y-p_y^{\prime})/\hbar\right]
={L_y\over{4a_B(2\pi a_Bk_z)^3}}e^{a_B^2k_z^2}\Phi(2Wk_z)
\end{eqnarray}
where
\begin{equation}
\Phi(\alpha)=\int_{1}^{\infty}
d\xi{e^{-\alpha\xi}\over{\xi^3\sqrt{\xi^2-1}}},
\end{equation}
For $\alpha\,\gg\,1$
\begin{equation}
\Phi(\alpha)\simeq\,\sqrt{{\pi\over{2\alpha}}}e^{-\alpha}.
\end{equation}
This result for the effective interaction (\ref{fin_form}) can
be explained as follows: Larmour circles within the quantum
wells of the width $L_x\cdot\,1/(k_zL_x)$ near the surfaces
contribute to the interaction. The number of interacting circles
from two quantum wells is $$
\left({L_y\over{a_B}}{L_x\cdot\,1/(k_zL_x)\over{a_B}}\right)^2.
$$ The sum over $q_x,q_x^{\prime}\sim\,k_z,
q_y\sim\,\sqrt{k_z/W}$ leads to a factor
$(k_zL_x)^2\cdot\,L_y\sqrt{k_z/W}$. The exponential decay of the
drag with the distance between the quantum wells $W$ is a
consequence of one-dimensional character of the drag in the
strong longitudinal magnetic field.  Combining all these factors
and multiplying the result by
$U^2\sim(4\pi\,e^2)^2/(L_xL_yL_z)^2k_z^4$ we arrive at
Eq.(\ref{fin_form}).

The product of the distribution functions in
Eq.~(\ref{drag_cur}) is a sharp function of $p_z$ and
$p_z^{\prime}$ at small temperatures, acquiring nonzero values
only at $p_z,p_z^{\prime}\sim\,p_F^B\pm\,T/v_F^B$. We assume
that the quasimomentum interval $T/v_F^B$ is much smaller than
$\hbar/W$
\begin{equation}\label{f_ineq}
T\,\ll\,{\hbar v_F^B\over{W}}.
\end{equation}

Here we wish to note that the Boltzmann treatment of transport
phenomena requires that the uncertainty in longitudinal momentum
must be smaller than the same momentum interval
$\hbar/L_z\,\ll\,T/v_F^B$. These two requirements automatically
lead to the inequality $W\,\ll\,L_z$. We assume that the last
inequality holds.

According to our assumptions we can regard
the interaction term in Eq.~(\ref{drag_cur}) as the slowly varying
function and obtain
\begin{eqnarray}\label{part_case_first}
J_{\mbox{drag}}=J_0
{eV\over4\varepsilon_F^B}{T\over\varepsilon_F^B}
\left({U_{12}\over{2T}}\right)^2
\left[\sinh\left({U_{12}\over{2T}}\right)\right]^{-2}
\end{eqnarray}
where
\begin{eqnarray}\label{part_cs01}
J_0=-
{e^5m\over{\kappa^2(4\pi\hbar)^3}}
{L_zL_y\over{a_B^2}}
{1\over{(a_Bk_F^B)^2}}
e^{(2a_Bk_F^B)^2}\Phi\left(4Wk_F^B\right)
\end{eqnarray}

Here we introduced notations $U_{12}=
U_{1}-U_{2}$ and
$mv_F^B=p_F^B=\sqrt{2m[\mu_B-U_{1}-\hbar\omega_B/2]}$, $k_F^B=p_F^B/\hbar$.

We assume that the electrons remain degenerate in the
magnetic field
\begin{equation}
\label{low_ineq}
\varepsilon_F^B\equiv\mu_B-U_1-{\hbar\omega_B\over{2}}\,\gg\,T.
\end{equation}

We consider the quantum limit, i.e. the case when all
electrons belong to the first Landau level
\begin{equation}\label{low_lan}
\varepsilon_F^B<\,\hbar\omega_B.
\end{equation}
Since the electron concentration $N_B$ under this condition is
related to the chemical potential by the equation
\begin{equation}
N_B={m\hbar\omega_Bp_F^B\over{\pi^2\hbar^3}}
\end{equation}
Eq.(\ref{low_ineq}) and Eq.(\ref{low_lan}) lead to
\begin{equation}\label{inequality}
T\,\ll\,{(p_F^B)^2\over{2m}}\,<\,\hbar\omega_B,\;\;p_F^B={\pi^2\hbar^3\over{m}}
{N_B\over{\hbar\omega_B}}.
\end{equation}
The first inequality in this relation is weaker than Eq.(\ref{f_ineq}) if
$\varepsilon_F\,\sim\,\hbar\omega_B$ and
$Wk_F\,\ge\,1\,$.
Introducing the electron concentration $N$ and the chemical potential
$\mu$ for $B=0$ given by
\begin{equation}
N={(2m\varepsilon_F)^{3/2}\over{3\pi^2\hbar^3}},\;\;\varepsilon_F=\mu-U_1
\end{equation}
one can rewrite Eq.(\ref{inequality}) as
\begin{equation}
T\,\ll\,{4\over{9}}\left({N_B\over{N}}\right)^2
\left({\varepsilon_F\over{\hbar\omega_B}}\right)^2\varepsilon_F\,
<\,\hbar\omega_B.
\end{equation}
Note that the second inequality in this expression does not
depend on the electron mass and can require magnetic fields
stronger than Eq.(\ref{ineq_10}) [thus imposing a constraint on
the electron concentration, or, if the latter is given the
inequality may require stronger magnetic fields than is
required by the Eq.(\ref{ineq_10})].  For instance, in a
magnetic field of the order of $B\sim\,10\,$T the electron
concentration $N$ must be smaller than
$2.7\cdot\,10^{17}\,\,$cm$^{-3}$.

Considering the case of the aligned quantum wells, so that
$U_1=U_2$ [otherwise the effect is exponentially small, cf.
Eq.(\ref{part_case_first})] and putting $N=N_B$ we obtain
\begin{eqnarray}\label{part_case_sec}
J_{\mbox{drag}}=J_0
{eV\over4T}\left({T\over\varepsilon_F}\right)^2
\left(3\hbar\omega_B\over{2\varepsilon_F}\right)^4,
\end{eqnarray}
\begin{eqnarray}\label{part_cs}
J_0=-
{e^5mL_yL_zk_F^2\over{9\kappa^2(4\pi\hbar)^3}}
\left({3\hbar\omega_B\over{2\varepsilon_F}}\right)^4
e^{12(2\varepsilon_F/3\hbar\omega_B)^3}
\Phi\left(4Wk_F{2\varepsilon_F\over{3\hbar\omega_B}}\right).
\end{eqnarray}
The drag current is a rapidly increasing function of the applied
magnetic field, as the latter increases the density of states and
decreases the transferred Fermi momentum.
\begin{figure}[htb]
\epsfxsize=5.4in
\epsffile{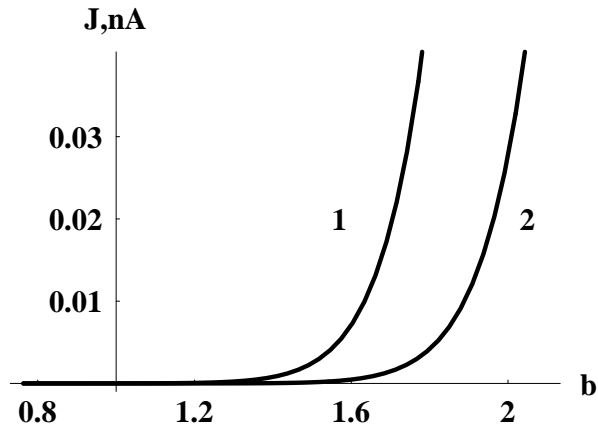}
\caption{Drag current versus dimensionless magnetic field
$b=\hbar\omega_B/\varepsilon_F$ for two values of the interwell
distances $W=40$nm (1) and $W=50$nm (2).
Other parameters are given in the text.}
\label{fig2}
\end{figure}

To make an estimate of the current we put $m=0.07m_e$,
$\hbar\omega_B\sim\varepsilon_F=14$meV, $\kappa=13$,
$L_z\sim\,L_y=1\mu$m, $W=40$nm.
$$
J_{\mbox{drag}}\sim\,10^{-11}\;\mbox{A}
$$

In the linear response regime
we can introduce a drag resistance, i.e. we can introduce the
coefficient that depends only on the quantum wells parameters and
relates the drive current $J_{\mbox{drive}}$ in the quantum well 2 to the induced
voltage in the drag quantum well $J_{\mbox{drive}}R_{\mbox{D}}=V_{\mbox{d}}$.
Here the drive current in
the quantum well 2 is
\begin{equation}
J_{\mbox{drive}}=-V{e^2\over{2\pi\hbar}}{L_xL_y\over{\pi\,a_B^2}}
\end{equation}
(cf. Eq.(\ref{dir_c}))
and $V_{\mbox{d}}$ is
determined by the condition of zero total current
$J=J_{\mbox{drag}}+J_{\mbox{Ohm}}=0$ in the drag
quantum well in Eq.(\ref{current}).
\begin{equation}\label{R_D}
R_{\mbox{D}}={\pi\hbar\over{e^2}}
{E_B\over{\varepsilon_F}}{T\over{\varepsilon_F}}{L_z\over{L_y}}
{1\over{(k_FL_x)^2}}
\left({3\hbar\omega_B\over{4\varepsilon_F}}\right)^6
e^{12(2\varepsilon_F/3\hbar\omega_B)^3}
\Phi\left[4Wk_F{2\over{3}}{\varepsilon_F\over{\hbar\omega_B}}\right],
\end{equation}
where we introduced effective Bohr energy 
$E_B={me^4/{\kappa^2\hbar^2}}$
and
\begin{equation}
p_F^B={2\over{3}}{N_B\over{N}}{\varepsilon_F\over{\hbar\omega_B}}p_F,\;\;p_F=
\sqrt{2m\varepsilon_F}.
\end{equation}
With the given above parameters we have the following estimate for
the transresistance
$$
R_{\mbox{D}}\sim\,0.4\;\mbox{m}\Omega.
$$
Now let us discuss when one can neglect the screening. Since the
transferred momenta are $q_z\sim\,2p_F^B/\hbar$ we may not take into account
the screening of the Coulomb potential if the inverse screening length
is much smaller than the transferred momentum.
We estimate the screening length at a transferred energy $\sim\,T$ as
\begin{equation}\label{screen01}
{1\over{r_s}}\sim\sqrt{{\pi e^2N_B\over{\kappa \varepsilon_F^B}}
\ln{{\varepsilon_F^B\over{T}}}}.
\end{equation}
The required inequality can be written as (we put $N_B=N$)
\begin{equation}\label{screen02}
{N^{1/3}e^2\over{\kappa}}
\ln{\left[{\varepsilon_F\over{T}}
\left({\varepsilon_F\over{\hbar\omega_B}}\right)^2\right]}
\ll\,
\varepsilon_F
\left({\varepsilon_F\over{\hbar\omega_B}}\right)^4.
\end{equation}
We will assume this inequality to be satisfied.

\section{Non-ohmic case}\label{nohm}
The product of distribution functions
Eq.(\ref{ct_1}) can be recast into the form
\begin{eqnarray}\label{ct_2}
{\cal
P}=2\sinh{\left({eV/{2T}}\right)}
\exp\{(\varepsilon^{(1)}_{p_z}-\mu_B)/T\}
\exp\{(\varepsilon^{(2)}_{p_z\prime}-\mu_B)/T\}\\
\times f(\varepsilon^{(1)}_{p_z}-\mu_B)
f(\varepsilon^{(2)}_{p_z^{\prime}}-\mu_B-eV/2)\nonumber
f(\varepsilon^{(1)}_{p_z^{\prime}}-\mu_B)
f(\varepsilon^{(2)}_{p_z}-\mu_B+eV/2)
\end{eqnarray}

As $\cal P$ is a sharp function of $p_z$ and $p_z^{\prime}$ one
can take out of the integral all the slowly varying
functions and get
\begin{eqnarray}\label{nonlin}
\int_0^{\infty}{dp_z}
{dp_z^{\prime}}{{\cal P}\over{(p_z+p_z^{\prime})}}
\int_{-\hbar\,L_x/2a_B^2}^{\hbar\,L_x/2a_B^2}{2L_ydp_y\over{2\pi\hbar}}
{dp_y^{\prime}\over{2\pi\hbar}}
g_{00}\left[(p_z+p_z^{\prime})/\hbar,(p_y-p_y^{\prime})/\hbar\right]
\nonumber\\
={L_ym^2a_B^2T^2\exp{(2a_Bk_F^B)^2}\over{4(4\pi\hbar)^3(a_Bk_F^B)^6}}
\Phi(4Wk_F^B)
\sinh\left({eV\over2T}\right)
\frac{\displaystyle{eV\over4T}-{U_{12}\over 2T}}
{\sinh\left(\displaystyle{eV\over4T}-{U_{12}\over
2T}\right)}
\cdot\frac{\displaystyle{eV\over4T}+{U_{12}
\over 2T}}
{\sinh\left(\displaystyle{eV\over4T}+{U_{12}\over
2T}\right)}
\end{eqnarray}
The drag current is
\begin{eqnarray}\label{part_case}
J_{\mbox{drag}}=J_0{1\over2}\left({T\over{\varepsilon_F^B}}\right)^2\sinh\left({eV\over2T}\right)
\frac{\displaystyle{eV\over4T}-{U_{12}\over 2T}}
{\sinh\left(\displaystyle{eV\over4T}-{U_{12}\over
2T}\right)}
\cdot\frac{\displaystyle{eV\over4T}+{U_{12}
\over 2T}}
{\sinh\left(\displaystyle{eV\over4T}+{U_{12}\over
2T}\right)}
\end{eqnarray}
For $eV\ll T$ one gets from Eq. (\ref{part_case}) the result of
Eq.~(\ref{part_case_sec}). Let us consider the opposite case
$eV\gg T$.
In this case one gets a nonvanishing result for
Eq.(\ref{part_case}) only if
$|U_{12}|\,<\,eV/2$
and one obtains the following equation for the drag current
\begin{equation}
J_{\mbox{drag}}=J_0
\left[\left({eV\over4\varepsilon_F}\right)^2-
\left({U_{12}\over2\varepsilon_F}\right)^2\right]
\left(3\hbar\omega_B\over{2\varepsilon_F}\right)^4
\label{2}
\end{equation}
Thus the drag current vanishes unless $eV\,>\,2|U_{12}|$.
\section{Summary}
We have developed a theory of the Coulomb drag between two
quantum wells in a strong longitudinal magnetic field. We
have considered a comparatively simple limiting case where
only the lowest Landau level is occupied. The strong magnetic
field makes transverse motion of an electron one dimensional.
These one dimensional electron states can be visualized as quantum
"tubes" or "wires".  Therefore, the Coulomb drag problem in this
situation becomes similar to the Coulomb drag problem between
two parallel nanowires.

It is interesting to compare our results with two different
geometries of experiment. First, let us consider the influence
of magnetic field on 1D Coulomb drag for the longitudinal
geometry. In this case the magnetic field is directed along $z$
axis and is parallel to 1D nanowires. For simplicity, we assume that the confining
potential in the absence of the magnetic field is
\begin{equation}\label{pot}
U(x,y)={m\Omega^2\over{2}}(x^2+y^2).
\end{equation}
The applied magnetic field shortens the radius of the state $a_B$ so
that it becomes
\begin{equation}\label{loc_length}
a_B^2={a_0^2\over{\sqrt{1+\left(B/B_c\right)^2}}},\;\;B_c=2{\Omega
mc\over{|e|}},\;\;a_0=\sqrt{{\hbar\over{2m\Omega}}}
\end{equation}
where $a_0$ is the radius in the absence of the magnetic field.
For the lowest Landau level we have
\begin{equation}\label{wave_function}
\phi={1\over{\sqrt{2\pi}}}{1\over{a_B}}\exp{\left(-\rho^2/4a_B^2\right)},\;\;
\varepsilon_p={\hbar^2\over{2ma_B^2}}+{p_z^2\over{2m}}.
\end{equation}
The wave function of the electron in the second wire can be
obtained by a gauge transformation of the wave function in the
first one. Since the interaction term $g$ is not phase sensitive
we are left only with a shift by the distance $W$ between the
centers of the wires in the argument of the wave function
(\ref{wave_function}).  As a result, one gets for the interaction
\begin{equation}\label{g00}
g(2p_F^B)=4e^{-W^2/2a_B^2}\left[\int_0^{\infty} d\rho \rho e^{-\rho^2}
I_0\left({W\over{a_B}}\rho\right)K_0\left(4{p_F^Ba_B\over{\hbar}}\rho\right)
\right]^2,
\end{equation}
where $I_0(x),\,K_0(x)$ are the modified Bessel functions. The
quasimomentum
$$
p_F^B={1\over{2}}\pi\hbar N_L^B
$$
must satisfy the inequality
\begin{equation}\label{p0}
T\ll\,(p_F^B)^2/2m\,<\,{\hbar^2\over{2ma_B^2}},
\end{equation}
since we have assumed that only the lowest Landau level is occupied.
Here $N_L^B$ is the electron density per unit length in magnetic field.
The expression (\ref{g00}) demonstrates that provided the magnetic field
goes up
the localization radius $a_B$ of the wave functions suppresses
the probability of the backscattering processes.
Note that if one assumes $N_L^B=N_L$, where $N_L$
is the electron density per unit length for $B=0$ then
the effective interaction depends on the magnetic field only via
$a_B$. Therefore in this case the magnetic field does not change the
magnitude of transferred momentum, in contrast with the previous case where
such a change leads to a rapid increase of the drag current in a strong
magnetic field.
The drag current is
\begin{eqnarray}\label{part_1D}
J_{\mbox{drag}}=J_{01}
{eV\over T}\left({T\over\varepsilon_F^B}\right)^2
\left({U_{12}\over{2T}}\right)^2
\left[\sinh\left({U_{12}\over{2T}}\right)\right]^{-2}
\end{eqnarray}
\begin{eqnarray}\label{part_cs02}
J_{01}=-
{e^5m\over{2\pi^2\kappa^2\hbar^3}}
{L_zk_F^B}\,g(2p_F^B).
\end{eqnarray}

Second, we can compare our results with the drag between two
two-dimensional quantum wells~\cite{Gramila} in the field-free
case. In this case the transresistance $\rho_{12}$ is
proportional to
\begin{equation}
\rho_{12}\sim\,T^2 {1\over{(k_Sd)^2}}{1\over{(k_Fd)^2}},
\end{equation}
where $k_{S}$ is the single-quantum well (two-dimensional)
Thomas-Fermi screening wave vector of the order of the inverse
effective Bohr radius, $d$ is the interwell distance.
First we note that the temperature dependence of the Coulomb
drag between two (three-dimensional) quantum wells in the strong
magnetic fields is weaker than for the drag in two dimensions.
Second, we note that in the latter case the contribution from
the backscattering processes can be neglected as compared to the
small angle scattering contribution with transferred momenta
$0<q<\,1/d\ll\,k_{S}$ while in our case only the backscattering
processes are important (this is again a consequence of
one-dimensionality of the Coulomb drag problem in the quantum
limit).

\acknowledgements

The authors acknowledge support for this work by the Russian
National Fund of Fundamental Research (Grant No~03-02-17638).



\begin{references}
\bibitem{Bonsager} M. C. Bonsager, K. Flensberg, B. Y.-K. Hu and
A.-P. Jauho, Phys. Rev. Lett. {\bf 77}, 1366 (1996),\\
M. C. Bonsager, K. Flensberg, B. Y.-K. Hu and
A.-P. Jauho, Phys. Rev. B {\bf 56}, 10314 (1997).
\bibitem{Kamenev} A. Kamenev and Y. Oreg, Phys. Rev. B {\bf 52}, 7516,
(1995).
\bibitem{Hu} B. Y. K. Hu, Phys. Scripta {\bf T69}, 170 (1997).
\bibitem{Bogachek} E. N. Bogachek, M. Jonson, R. I. Shekhter and
T. Swahn, Phys. Rev. B {\bf 50}, 18341 (1994).
\bibitem{GurMur} V. L. Gurevich, V. B. Pevzner, and E. W. Fenton J.Phys.:
{\it Condens. Matter} {\bf 10}, 2551, (1998).
\bibitem{GM}V. L. Gurevich, M. I.  Muradov, Pis'ma v ZhETF, {\bf 71},
164, (2000) [JETP Letters {\bf 71}, 111 (2000)].
\bibitem{Lan}R. Landauer, IBM J. Res. Develop. {\bf 1}, 233 (1957);
{\bf 32}(3), 306 (1989).
\bibitem{LB}Y. Imry , Directions in Condensed Matter Physics, ed.
G. Grinstein and G. Mazenko, 1986 (Singapore: World Scientific),
101.
\bibitem{B}B\"uttiker M., Phys.Rev. Lett., {\bf 57}, 1761 (1986).
\bibitem{Gramila} T. J. Gramila, J. P. Eisenstein, A. H. MacDonald,
L. N. Pfeiffer and K. W. West, Phys. Rev. Lett., {\bf 66}, 1216 (1991).
\end{references}
\end{document}